\documentclass[a4paper, 12pt]{article} % use larger type; default would be 10pt

\usepackage[utf8]{inputenc} % set input encoding (not needed with XeLaTeX)

\usepackage{geometry} % to change the page dimensions
\usepackage{graphicx} % support the \includegraphics command and options

\usepackage[parfill]{parskip} % Activate to begin paragraphs with an empty line rather than an indent

\usepackage{amsmath} % for tfrac
\usepackage{booktabs} % for much better looking tables
\usepackage{array} % for better arrays (eg matrices) in maths
\usepackage{paralist} % very flexible & customisable lists (eg. enumerate/itemize, etc.)
\usepackage{verbatim} % adds environment for commenting out blocks of text & for better verbatim
\usepackage{subfig} % make it possible to include more than one captioned figure/table in a single float
\usepackage{multicol}
\usepackage{sectsty}
\allsectionsfont{\sffamily\mdseries\upshape} % (See the fntguide.pdf for font help)
\usepackage[nottoc,notlof,notlot]{tocbibind} % Put the bibliography in the ToC
\usepackage[titles,subfigure]{tocloft} % Alter the style of the Table of Contents

\usepackage{epstopdf}

\title{Two approximate screening functions \\ for the neutral Thomas-Fermi atom}
\author{Herbert E. Müller \\ \normalsize http://herbert-mueller.info/}
\date{} % Activate to display a given date or no date (if empty),
\begin{document}
\maketitle

\begin{flushleft}

\begin{abstract}
In the Thomas-Fermi model for the atom or ion, the electric potential $\phi$ and electron density $n$ are both expressed in terms of a screening number $\chi$ and the radius  $x$. The second order differential equation for the screening function $\chi(x)$ cannot be solved in closed form. Two approximate formulas for the screening function of the neutral atom are proposed: the first formula is analytically simple, the second formula has the roles of dependent  and independent variable inter- changed. Both formulas  reproduce the true screening function very accurately. 
\end{abstract}

\section{Introduction}

In quantum chemistry, electron densities of atoms and molecules are described with the Thomas-Fermi model and refinements thereof. The second order differential equations for the electric potential that come with these models cannot be solved in closed form. The approach is then to find approximate multi-parameter solutions, and to adjust the parameters such that the electronic binding energy becomes minimal. In the Thomas-Fermi model for the single atom (or ion) the electric potential $\phi$ is set $\phi=\chi/x$, with $\chi$ the screening number and  $x$ the radius. Finding approximate screening functions $\chi(x)$ is an open-ended guessing game, that (in the case of the neutral Thomas-Fermi atom at least) has lead to too many publications already. So why do I trouble you with my two proposals? Well, as to the first formula, given its analytical simplicity, the numerical precision is astonishing. This formula practically replaces a numerical table. The second formula has the roles of dependent  and independent variable interchanged, i. e. it expresses $x$ as a function of $\chi$. It is a bit more complex, but still numerically very precise. Expressing $x$ in terms of $\chi$ instead of the other way round may be fruitful idea for the description of the Thomas-Fermi ion, too. 

\pagebreak
\section{Notation, basic formulas}

The theory of the Thomas-Fermi atom or ion is covered in most elemantary textbooks on quantum mechanics. The notation and basic formulas are therefore given below without further explanation.\\
\quad 

\begin{multicols}{2}
Unit of Length: \(0.8853 Z^{-1/3}\cdot a_B\)

Radius from the nucleus: \textit{x} 

Nuclear Potential (\textit{p} only):  \(-\frac{1}{x}\)

Unit of Energy: \(2.2590 Z^{4/3}\cdot Ry\) 

Ion radius: \textit{X} 

Overall Potential (\textit{p} and \textit{e}):  \( \phi = -\frac{\chi}{x}-b\)
\end{multicols}
Screening function: $\chi$, with $\chi(0)=1$ and $\chi(X)=0$

Slope of the screening function: $\psi=-\frac{d\chi}{dx}$, with $\psi(0)=a$ and $\psi(X)=b$

Fraction of electrons inside radius \textit{x}: \(N=-x^2\frac{d}{dx}\frac{1-\chi}{x}\), with  \(N(X)\leq1\)

Electron number density: \(n=-\frac{1}{x^2}\frac{dN}{dx}=\frac{1}{x}\frac{d^2\chi}{dx^2}\)

Overall Binding energy:  \( B=\int_0^\infty dx x^2n \left( \phi-\frac{3}{5}n^{2/3} \right)  \)

For the neutral atom, $X=\infty$, $b=0$, and $N(X)=1$.

\section{Thomas-Fermi differential equation, series expansions}

The Thomas-Fermi differential equation for the screening function $\chi(x)$ is:

\begin{equation}
\frac{d^2\chi}{dx^2} = \frac{\chi ^{3/2}}{x^{1/2}}
\end{equation}

The boundary conditions for the screening function of the neutral TF-atom are  
\begin{equation}
\chi(0)=1 \qquad  \qquad \forall x>0: \ \chi>0  \qquad  \qquad  \chi \rightarrow 0  \quad (x  \rightarrow \infty)
\end{equation}

The Thomas-Fermi differential eqn. cannot be solved in closed form. 

The Taylor-series at $x=0$ for the screening function \(\chi(x)\) of the Thomas-Fermi ion (including the neutral atom)  was given by Baker \cite{Englert}: 
\begin{equation}
\chi=1-ax+\frac{4}{3}x^{3/2}-\frac{2}{5}ax^{5/2}+\frac{1}{3}x^3 \ldots 
\end{equation} 
with $a\leq1.588072$ (equality for the neutral atom). 
  
For the neutral Atom, the Taylor series of $\chi(x)$ at \(x=\infty\) was given by Coulson and March \cite{Englert}: 
\begin{equation}
\chi=\frac{144}{x^3}\left(1-Fx^\lambda+0.6256(Fx^\lambda)^2-0.3133(Fx^\lambda)^3 \ldots \right)
\end{equation}
with \(F= 13.270\) and \(\lambda=-0.7720\).

Both series converge very slowly. 

\section{First approximate screening function}

The following formula interpolates between the Taylor expansions of $\chi(x)$ at \(x=0\)  and at \(x=\infty\) :
\begin{equation} 
\chi=\left( 1+\frac{ax}{\alpha}\left( 1+\frac{4\sqrt{x}}{3\beta a}\right)^{-\beta} \right)^{-\alpha}
\end{equation}
This formula has three free parameters: the initial slope of the the screening function \(a=\left(-d\chi/dx\right)_0\), and the two exponents \(\alpha\) and \(\beta\). These can be fixed by comparison with the Baker series  and the asymptotic formula. Formula (5) reproduces the Baker series up to the power \(x^{3/2}\). The condition that the Taylor coefficient of \(x^2\) vanishes leads to 
\[\frac{1+1/\alpha}{1+1/\beta}=\frac{16}{9a^3}\]

The condition that \(\chi\) falls off as \(x^{-3}\) at large \(x\) leads to 
\[\alpha(1-\beta/2)=3\]

Combining these two eqn.s and resolving for the two exponents gives 
\begin{equation} 
\alpha=1+\sqrt{\frac{12a^3-8}{3a^3-8}}
\end{equation}
and
\begin{equation} 
\beta=2\left(1-\frac{3}{\alpha}\right) 
\end{equation}

Now we must somehow fix the value of  the parameter $a$.

If we want to describe the electronic density by \(n=(\chi/x)^{3/2}\) (which is more convenient than Poisson's equation $n=\chi''/x$), we can fix $a$ by the condition that the integrated electron density be 1: 
\[N(a,\alpha,\beta)\equiv\int_0^\infty dx x^{1/2}\chi^{3/2} =1 \]
In this way we obtain the parameter set 
\begin{equation} 
\begin{array}{l l l}
a=1.58968 & \quad \alpha=4.1501 & \quad \beta = 0.55426 
\end{array}
\end{equation}

The absolute error of $\chi$ in eq. (5) is now $<$3e$-4$ for all $x$, and the relative error grows to $4\%$ for $x \rightarrow \infty$. 

Alternatively, we may simply use the value of $a$ known from numerical integration. The corresponding parameter set is $a=1.58807$, $\alpha=4.1587$, $\beta = 0.55723$.  

Finally, if we want to reproduce the asymptotic behaviour $\chi\rightarrow 144/x^3$, we must set $a=1.58662$, $\alpha=4.1665$, $\beta = 0.55994$. 

These numbers are all very close to each other. By plotting the absolute error $\Delta\chi(x)$  for the three parameter sets, the author found that set (8) gives the "best" overall precision. The approximate screening function (5), (8) and the true screening function are compared graphically in section 7.

\section{Sommerfeld's screening function} 

Sommerfeld introduced the three parameter approximate screening function 
\begin{equation} 
\chi=(1+(kx)^{1/\beta})^{-\alpha}
\end{equation}
because of its analytical tractability (the exponent $1/\beta$ instead of $\beta$ will turn out to be more convenient). 

Sommerfeld chose the parameter values  $k=12^{-2/3}=0.1908$, $\alpha=3.886$, $\beta=\alpha/3$ $=1.295$, others modified the numbers. Here we will fix $k$, $\alpha$, $\beta$ by minimizing the energy functional $S[\chi]$. This is done as follows. 

The Thomas-Fermi differential eqn. (1) is the Euler-Lagrange equation
\begin{equation*} 
\frac{\partial L}{\partial \chi}=\frac{d}{dx} \frac{\partial L}{\partial \chi'} 
\end{equation*} 
to the Lagrangian   
\begin{equation} 
L=\frac{\chi'^2}{2}+\frac{2}{5}\frac{\chi^{5/2}}{x^{1/2}}
\end{equation}
The prime indicates $d/dx$. 

The minimal action to this Lagrangian, evaluated with the $\chi(x)$ satisfying the TF differential eqn. (1), is 
\begin{equation} 
S[\chi(x)]=\int_0^\infty dxL(\chi,\chi',x)=\frac{3}{7}a=B
\end{equation}

In words: the minimal action is 3/7th of the initial slope $a$ of $\chi(x)$, and equals the electronic binding energy $B$. 

The action evaluated with Sommerfeld's trial function is 
\begin{equation} 
S(k,\alpha,\beta)=k\frac{\alpha^2}{2\beta}B\left(2-\beta,2\alpha+\beta\right)+ k^{-1/2}\frac{2\beta}{5}B\left(\frac{\beta}{2},\frac{5\alpha-\beta}{2}\right)
\end{equation}

A numerical analysis shows the action is minimal for the parameter values 
\begin{equation} 
\begin{array}{l l l}
k\approx0.482 & \quad \alpha\approx2.10 & \quad \beta\approx1.083 
\end{array}
\end{equation} 

The minimal action gives an upper limit for the binding energy $B$ and the initial slope $a$ of the ``true'' screening function, see eqn. (11). 
\begin{equation}
\begin{array}{l l }
S_{\min}=0.68075>B(=0.68060) & \quad \frac{7}{3}S_{\min}=1.58841>a(=1.58807)
\end{array}
\end{equation}

With the parameter set  (13), the absolute error of $\chi$ in eqn. (9) is $<$7e$-3$ for all $x$, while the relative error grows infinitely for $x \rightarrow \infty$. 

\pagebreak
\section{Second approximate screening function}

The procedure outlined above can be transfered to other quantities such as the inte- grated electron number $N(x)$, or the electron number density $n(x)$: one tries to describe it with the 4-parameter expression \(x^{\gamma}(1+(kx)^{1/\beta})^{-\alpha}\), which can be manipulated relatively easily, finds the Lagrangian and action to this quantity, and then fixes $\alpha, \beta, \gamma$ and $k$ by minimizing or maximizing the action. Actually, doing exactly this leads not to significant improvement in the quality of the approximate formulas for $\chi(x)$. We need some new idea. And here it is: 

\subsection{Use $\chi$ as the independent variable!}

The slope of the screening function is
\begin{equation}      
 \psi \equiv -\frac{d\chi}{dx}
\end{equation} 
For the neutral TF-atom, 
\begin{equation}
\psi(0)=a \qquad  \qquad \psi>0 \quad (x>0)   \qquad  \qquad  \psi \rightarrow 0  \quad (x  \rightarrow \infty)
\end{equation}
%(For the Thomas-Fermi ion, $\psi(X)=b$: Fermi-energy.)

We now use $\chi$ as the independent variable, and try to fit $\psi(\chi)$ for the neutral atom with the trial function 
\begin{equation}      
\psi=a(1-\bar\chi^{1/\beta})^{\alpha} 
\end{equation} 
where
\begin{equation}      
\bar\chi=1-\chi
\end{equation} 
Note the formal similarity of (17) with Sommerfeld's Ansatz (9)!

The series formulas $\psi=a-2\sqrt{\bar\chi/a}+\ldots$ ($\bar\chi\approx 0$) and $\psi = \chi^{4/3}/12^{2/3}+ \ldots$ ($\chi\approx 0$) suggest that $\beta\approx2$ and $\alpha\approx4/3$. 

The Ansatz (17) together with (15) leads to the inverse screening function  
\begin{eqnarray} 
x&=&\int^1 \frac{d\chi}{\psi} = \frac{1}{a}\int_0 d\bar\chi (1-\bar\chi^{1/\beta})^{-\alpha} \nonumber \\ 
&=& \frac{\beta}{a}B({\bar\chi^{1/\beta},\beta,1-\alpha})  \\
&=&\frac{\bar\chi}{a}F(\beta,\alpha,\beta+1,{\bar\chi^{1/\beta}}) \nonumber
\end{eqnarray} 

Here, $B$ is the incomplete Beta function, and $F$ is the hypergeometric function \cite{Arfken}. \\ (Unfortunately, the conventions for the order of the arguments in the incomplete Beta function and in the hypergeometric function disagree with each other.)  

Now let us calculate the best values of the free parameters $a$, $\alpha$ and $\beta$.  

\pagebreak
\subsection{The Lagrangian of $\eta(\chi)$}

The differential equation for $\psi(\chi)$ can be derived from the TF differential eqn. (1). It is more easily expressed in terms of  
\begin{equation}      
\eta=\psi^2 
\end{equation} 
and turns out to be 
\begin{equation} 
\frac{1}{2\sqrt\eta}=\frac{d}{d\chi} \frac{-2\chi^3}{\eta'^2} 
\end{equation} 
where the prime denotes now $d/d\chi$. 

Equation (21) is the Euler-Lagrange equation   
\begin{equation*} 
\frac{\partial L}{\partial \eta}=\frac{d}{d\chi} \frac{\partial L}{\partial \eta'} 
\end{equation*} 
to the Lagrangian 
\begin{equation} 
L=\sqrt{\eta}+2\frac{\chi^3}{\eta'} 
\end{equation} 

The action of this Lagrangian is 
\begin{equation} 
S[\eta(\chi)]=\int_0^1 d\chi L(\eta,\eta';\chi)
\end{equation} 

Let us calculate the action $S=S_1+S_2$ we get for the $\eta(\chi)$ satisfying eqn. (21). We have
\begin{eqnarray*} 
S_1&=&\int_0^1 d\chi \sqrt{\eta}=\int_0^1 d\chi \psi  \\
S_2&=&2\int_0^1 d\chi \frac{\chi^3}{d\eta/d\chi}=\int_0^1 d\chi \frac{\chi^3}{-d\psi/dx}=
\frac{1}{5}\int_0^1 dx \chi^{3/2}\sqrt{x}=\frac{1}{5} \int_0^1 dx \frac{\chi^{3/2}}{\sqrt{x}}
\end{eqnarray*} 

According to \cite{Englert} these integrals are (for the neutral atom) $S_1=2a/7$ and $S_2= a/7$. Thus we obtain again 
\begin{equation} 
S[\eta(\chi)]=\frac{3}{7}a=B
\end{equation} 

\subsection{The parameter $a$}

We would now like to evaluate the action (23) with the trial function (17) and eqn. (20). However there seems to be a problem: what value should we choose for $\psi(\chi=1) \equiv a$? From numerical integration we know its value to be 1.588072, but it seems unsatisfactory that we have to use this information. \\In variational problems, the boundary values of the function to be determined may normally not be varied. Fortunately, we have here an exception to this rule: the variation of the action is  
\begin{equation*} 
\delta S=\frac{\partial L}{\partial \eta'}
|_0^1+\int_0^1 d\chi \delta \eta
\left(\frac{\partial L}{\partial \eta}-\frac{d}{d\chi} \frac{\partial L}{\partial \eta'}  \right) 
\end{equation*} 

For our trial function (17), $\partial L/\partial \eta'$ vanishes at $\chi=1$ if $\beta>1$ (which is what we expect), i. e. $a$ may be varied along with  $\beta$ and $\alpha$ !  

\pagebreak
\subsection{The action $S[\eta(\chi)]$}

To evaluate of the action (23) we set
\begin{equation} 
1-\chi=t^\beta 
\end{equation} 
The action is of the form 
\begin{equation} 
S(a,\alpha,\beta)=aI_1(\alpha,\beta)+a^{-2}I_2(\alpha,\beta)
\end{equation} 

The first integrals $I_1$ is
\begin{equation} 
I_1(\alpha,\beta)=\int_0^1 d\chi \frac{\psi}{a}=\beta\int_0^1 dt t^{\beta-1}(1-t)^\alpha =
\beta B(\beta,\alpha+1) 
\end{equation} 
The second integrals $I_2$ is
\begin{equation*} 
I_2(\alpha,\beta)=2\int_0^1 d\chi \frac{a^2\chi^3}{d\eta/d\chi}=\frac{\beta^2}{\alpha}\int_0^1 dt
\frac{t^{2(\beta-1)}(1-t^\beta)^3}{(1-t)^{2\alpha-1}} \nonumber 
\end{equation*} 
We integrate $I_2$ by parts: $(1-t)^{1-2\alpha}$ is integrated, and $t^{2(\beta-1)}(1-t^\beta)^3$ is differentiated. This gives 
\begin{eqnarray} 
I_2(\alpha,\beta)&=&\frac{\beta^2}{2\alpha(\alpha-1)}\int_0^1 dt (1-t)^{2-2\alpha}\left[-(2\beta-2)t^{2\beta-3} \ldots \right. \nonumber \\
&&\left. +3(3\beta-2)t^{3\beta-3}-3(4\beta-2)t^{4\beta-3}
+(5\beta-2)t^{5\beta-3} \right] \nonumber \\
&=& \frac{\beta^2\Gamma(3-2\alpha)}{2\alpha(\alpha-1)}
\left[-\frac{\Gamma(2\beta-1)}{\Gamma(2\beta+1-2\alpha)}+\frac{3\Gamma(3\beta-1)}{\Gamma(3\beta+1-2\alpha)} \ldots  \right. \nonumber \\
&&\left. \qquad \qquad \quad \quad-\frac{3\Gamma(4\beta-1)}{\Gamma(4\beta+1-2\alpha)}
+\frac{\Gamma(5\beta-1)}{\Gamma(5\beta+1-2\alpha)} \right]
\end{eqnarray} 

Minimizing the action (26) with respect to $a$ gives
\begin{eqnarray} 
a^*(\alpha,\beta)&=&\sqrt[3]{2I_2/I_1} \\
S^*(\alpha,\beta) \equiv S(a^*,\alpha,\beta)&=&(2^{1/3}+2^{-2/3})I_1^{2/3}I_2^{1/3}
\end{eqnarray} 

The values of $\alpha, \beta$ that minimize $S^*(\alpha,\beta)$ can be read off a contour plot of $S^*(\alpha,\beta)$. Once they are found, eqn. (29) yields $a$. These numbers are 
\begin{equation}
\begin{array}{l l l }
\alpha = 1.382 & \quad \beta = 1.644  & \quad a = 1.550
\end{array}
\end{equation}

From eqn.s (30) and (11) we now get
\begin{equation}
\begin{array}{l l }
S_{min} = 0.68063>B(=0.68060) &  \quad \frac{7}{3}S_{min}= 1.58814 >a(=1.58807)
\end{array}
\end{equation}

With the parameter set  (31), the absolute error of $\chi$ in eq. (19)  is $<$6e$-4$ for all $x$, while the relative error grows infinitely for very large $x$. 

\pagebreak
\section{Comparison}

In figure (1) we compare Sommerfeld's fit (9), (13), the first proposed fit (5),  (8), and  the second proposed fit (19), (31), with the numerically integrated screening function. Both new fits are practically indistinguishable from the "true" screening function. 

\begin{figure}[!h]
\begin{center}
\includegraphics[clip = true, trim = 5pt 0pt 5pt 0pt, scale=1]{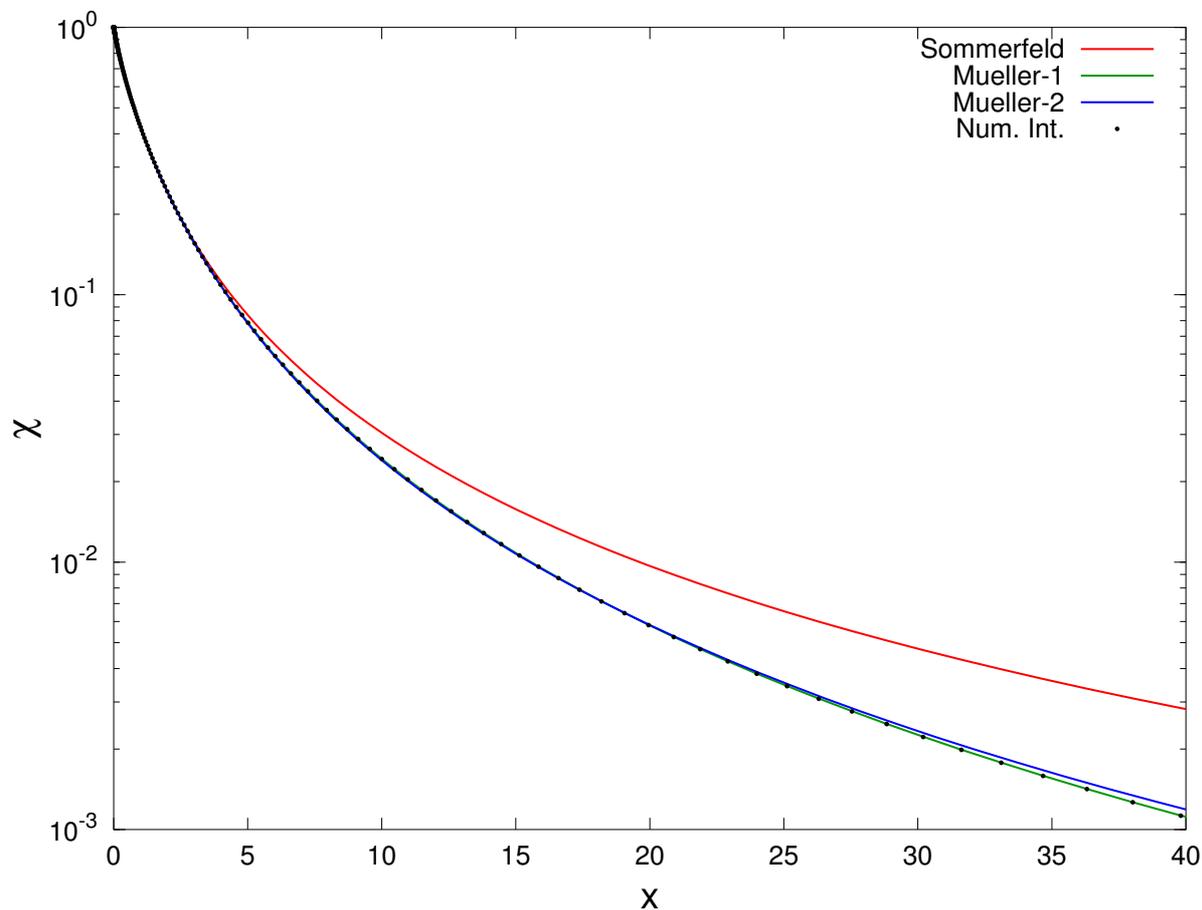}
\caption{Screening function: exact (dots) and fits (solid lines)}
\end{center}
\end{figure}

Let us also summarize the obtained precision in $\chi(x)$: 

\begin{tabular}{l l l l}
Fit & eqn.s & abs. error $\Delta\chi$ & rel. error $\Delta\chi/\chi$  \\
\hline
Sommerfeld & (9), (13) & $<$7e$-3$ & $<\infty$ \\
M\"uller-1 & (5), (8) & $<$3e$-4$ & $<4\%$ \\
M\"uller-2 & (19), (31) & $<$6e$-4$ & $<\infty$ \\
\end{tabular}

\end{flushleft}
\end{document}